\documentclass[useAMS,usenatbib]{mn2e}

\usepackage[dvips]{graphicx}
\usepackage{amsmath,amssymb}
\usepackage{color}

\title[Short-term evolution of barred grand-design spirals]
{Short-term dynamical evolution of grand-design spirals in barred galaxies}
\author[J. Baba]{Junichi \textsc{Baba}\thanks{E-mail:babajn@elsi.jp; babajn2000@gmail.com}\\
Earth-Life Science Institute, Tokyo Institute of Technology, 2--12--1 Ookayama, Meguro, Tokyo 152--8551, Japan.
}

\begin{document}

\date{Accepted 2015 September 22.  Received 2015 September 22; in original form 2015 July 30}


\maketitle

\begin{abstract}
We investigate the short-term dynamical evolution of stellar grand-design spiral arms in barred spiral galaxies using a three-dimensional (3D) $N$-body/hydrodynamic simulation. Similar to previous numerical simulations of unbarred, multiple-arm spirals, we find that grand-design spiral arms in barred galaxies are not stationary, but rather dynamic. This means that the amplitudes, pitch angles, and rotational frequencies of the spiral arms are not constant, but change within a few hundred million years (i.e. the typical rotational period of a galaxy). We also find that the clear grand-design spirals in barred galaxies appear {\it only when} the spirals connect with the ends of the bar. Furthermore, we find that the short-term behaviour of spiral arms in the outer regions ($R>$ 1.5--2 bar radius) can be explained by the swing amplification theory and that the effects of the bar are not negligible in the inner regions ($R<$ 1.5--2 bar radius). These results suggest that, although grand-design spiral arms in barred galaxies are affected by the stellar bar, the grand-design spiral arms essentially originate not as bar-driven stationary density waves, but rather as self-excited dynamic patterns. We imply that a rigidly rotating grand-design spiral could not be a reasonable dynamical model for investigating gas flows and cloud formation even in barred spiral galaxies.
\end{abstract}

\section{Introduction}
\label{sec:intro}

Spiral arms in disc galaxies can be classified into three categories: grand-design, 
multiple-arm, and flocculent spirals \citep{ElmegreenElmegreen1987}.
Grand-design spiral arms are large-scale coherent, symmetric, two-armed patterns (e.g. NGC 628),
whose number fraction is more than $50\%$ in nearby spiral galaxies \citep{Grosbol+2004,Kendall+2011}.
Observations show that grand-design spirals are associated with bars or companions 
\citep{KormendyNorman1979,SeigarJames1998,Kendall+2011},
suggesting that a bar or companion is essential in forming a grand-design spiral in a disc galaxy.
In barred spiral galaxies, the grand-design spirals extend from the ends of the bars. 
Furthermore, although some studies found little or no correlation 
between bar strengths and spiral arm strengths \citep{Durbala+2009,Kendall+2011}, 
other studies have found correlations \citep{Block+2004,Salo+2010}.
These observations, therefore, suggest that bars drive grand-design spirals with the {\it same} pattern speed as the bars.

This `bar-driven spiral hypothesis' has been promoted by the ballistic closed-orbit theory,
which is based on non-self-gravitating hydrodynamic simulations in fixed barred potentials 
\citep[e.g.][]{SandersHuntley1976,Sanders1977,Huntley+1978,Sormani+2015}.
This theory explains the physical mechanism responsible for the formation of the spirals in barred galaxies 
in terms of closed orbits of the gas \citep[see also][]{Wada1994}: the spiral arms in barred galaxies are regarded as 
{\it kinematic} density waves (i.e. crowding of gaseous closed orbits) driven by an external barred potential.
However, spiral arms in observed barred galaxies are composed of not gas but stars; 
therefore, theories based on stellar dynamics are required for investigating the origin of spiral arms in barred galaxies.

On the other hand, the `invariant manifold theory' is another bar-driven spiral theory 
\citep[][]{Romero-Gomez+2006,Romero-Gomez+2007,Athanassoula+2009b,Athanassoula+2009a,Athanassoula+2010}
and has been developed from stellar orbital theories in fixed barred potentials \citep{Danby1965}.
Essentially, the back-bones of barred spirals are bunches of untrapped stars 
escaped from unstable Lagrangian points $L_1$ and $L_2$ close to the ends of the bar.
This means that stars should {\it move along} the arms rather than across the arms \citep{Athanassoula2012}.
This behaviour is completely different from the quasi-stationary\footnote{
In general, the word stationary or steady means `not moving.'
However, in this context, this word indicates that density waves do not propagate radially
and that instead they propagate azimuthally with a single pattern speed.
See \citet{BertinLin1996} for further discussion on the concept of quasi-stationary density waves.
} density wave theory, which predicts that stars should {\it traverse} the arms 
but would stay longer in the arm than in the inter-arm region \citep{LinShu1964}. 
Furthermore, the manifold theory predicts a rigidly rotating spiral arm with 
a similar pattern speed to that of the bar (i.e. $\Omega_{\rm spiral} \simeq \Omega_{\rm bar}$)
and that stronger bars should have more open spirals compared to weaker bars \citep{Athanassoula+2009a}.
Recently, \citet{Struck2015}  extended the classical epicyclic orbit approximation to 
the `p-ellipse' orbit approximation \citep{Struck2006} and proposed that ensembles of 
eccentric resonant orbits excited in Lindblad zones can provide a backbone for 
generating kinematic bars and spiral waves. Furthermore, \citet{Struck2015} showed 
that some resonant eccentric orbits have a similar appearance to invariant manifold orbits.

The manifold theory and eccentric resonant orbits theory, however, do not consider the self-gravity of stars,
despite the fact that the self-gravity of stellar discs plays an essential role in the dynamical evolution of spirals via 
the (non-linear) swing amplification mechanism
\citep{SellwoodCarlberg1984,CarlbergFreedman1985,Fujii+2011,Baba+2013,D'Onghia+2013,D'Onghia2015}.
The swing amplification mechanism, proposed by Toomre in \citeyear{Toomre1981}, operates through 
a combination of three aspects, i.e. the shearing flow, epicyclic motions, and the disc self-gravity, in disc galaxies\footnote{
The term `swing amplification' is used to express the amplification process, which combines the shearing flow, epicyclic motions, 
and disc self-gravity, regardless of whether the process is a linear or nonlinear phenomenon. 
Originally, the `swing amplification' and its feedback cycle were based on the linear perturbation analyses 
of local shearing sheets \citep{GoldreichLynden-Bell1965,JulianToomre1966,Toomre1981}, 
and previous $N$-body simulations of galactic discs partly supported this mechanism \citep{CarlbergFreedman1985,Fujii+2011}. 
However, $N$-body simulations of galactic discs suggested the importance of nonlinearity, 
such as mode-mode coupling and radial migration of stars \citep{Baba+2013,D'Onghia+2013}, 
in the dynamical evolution of spiral arms.}.
Because the amplification is most efficient around the corotating radius \citep[e.g.][]{Shu1992b}, 
prominent spiral arms tend to have differentially rotating patterns, 
which almost follow the galactic rotation at every radius \citep[$\Omega_{\rm spiral} \simeq \Omega$;][]{Baba+2013}.
Besides, the pitch angles are expected to depend on the shear rates of galactic discs 
\citep{Fuchs2001a,Baba+2013,MichikoshiKokubo2014}.

For investigating the origin of grand-design spirals in barred galaxies, $N$-body and $N$-body/hydrodynamic simulations are required, 
and several preliminary numerical studies have been proposed. 
However, the studies have not reached a consensus about the dynamics of barred spirals.
Some numerical simulations of barred spiral galaxies have shown that the bar and spiral arm are rigidly rotating patterns 
with {\it different} pattern speeds, and they are independent patterns \citep{SellwoodSparke1988,RautiainenSalo1999}
or are coupled via non-linear interactions \citep{MassetTagger1997,RautiainenSalo1999,Minchev+2012}.
Recently, \citet{Roca-Fabrega+2013} performed $N$-body simulations of pure stellar barred spiral galaxies
and analysed the spiral rotation frequency from {\it selected} snapshots where the spiral amplitude
is above 70\% of the maximum amplitude; they concluded that the spiral rotation frequency 
approaches the bar's rigid body rotation in barred galaxies with the increment of bar strength.
On the other hand, \citet{Baba+2009} performed $N$-body/hydrodynamic simulations of a barred spiral galaxy 
and suggested that the spiral arms are not rigidly rotating patterns 
but transient recurrent patterns (i.e. dynamic spirals), 
and these results are similar to those obtained from multiple arm spiral galaxies
\citep[][]{SellwoodCarlberg1984,Fujii+2011,Wada+2011,Grand+2012a,Baba+2013,D'Onghia+2013}.
\citet{Grand+2012b} also found that spiral arms in barred galaxies are dynamic patterns whose 
rotation frequencies decrease with the radius in such a way that the rotation frequency is similar to the rotation of stars.

In this study, to address the short-term (e.g. a few hundred million years) behaviour of grand-design spirals in barred spiral galaxies, 
we performed a three-dimensional (3D) $N$-body/hydrodynamic simulation of a Milky-Way-like barred spiral galaxy.
This paper is organized as follows. We describe the galaxy model and our numerical simulation methodologies in Section \ref{sec:method}. 
In Section \ref{sec:results}, we present our results concerning the short-term behaviour of grand-design spirals in the barred galaxy 
and effects of the bar on spiral dynamics. Finally, we summarize our results in Section \ref{sec:summary}.
Long-term (e.g. $\sim 10$ Gyr) behaviour of spirals in barred galaxies, 
as well as effects of a `live' dark matter (DM) halo, will be presented in a forthcoming paper (M. S. Fujii et al. in preparation).

\section{Numerical Simulations and Analysis}
\label{sec:method}
 
We performed a 3D $N$-body/hydrodynamic simulation of a Milky Way-like galaxy 
with an $N$-body/smoothed particle hydrodynamics (SPH) simulation code {\tt ASURA-2} \citep{SaitohMakino2009,SaitohMakino2010}.
In this study, we focus on the dynamical evolution of spiral arms in a simulated barred spiral galaxy,
while detailed studies on spatial distributions of the interstellar medium (ISM) around spiral arms 
as well as the structures of the ISM/clouds and their galactic environmental dependences are separately discussed in 
Baba, Morokuma-Matsui \& Egusa (\citeyear{Baba+2015a}) and Baba, Morokuma-Matsui \& Saitoh (\citeyear{Baba+2015b}), respectively.

\subsection{Numerical methods}
\label{sec:NumericalMethod}

The self-gravities of stars and SPH particles were calculated by Tree with the GRAvity PipE (GRAPE) method \citep{Makino1991}.
The Tree method allows us to reduce the number of particle pair interactions that must be computed by dividing the volume into cubic cells, 
so that particles in distant cells are treated as a single large particle centred at the cell's centre of mass. 
The GRAPE tool is a special-purpose hardware for calculating gravitational forces. 
In this study, we used a software emulator of GRAPE, known as Phantom-GRAPE \citep{Tanikawa+2013}. 

Radiative cooling and heating, star formation, and stellar feedback from 
Type II supernovae and ${\rm H_{II}}$ regions were also included. 
The radiative cooling of the gas was solved assuming an optically thin cooling function
for a wide temperature range of $20~{\rm K} < T < 10^8~{\rm K}$ \citep{Wada+2009}.
Star formation from cold, dense gas and feedback models was based on the simple stellar population (SSP) approximation 
with the Salpeter initial mass function. These were implemented in a probabilistic manner \citep{Saitoh+2008}.
The ${\rm H_{II}}$-region feedback was implemented using a Stromgren volume approach \citep{Hopkins+2012,Baba+2015b}.

\subsection{Galaxy model}
\label{sec:GalaxyModel}

\begin{figure}
\begin{center}
\includegraphics[width=0.45\textwidth]{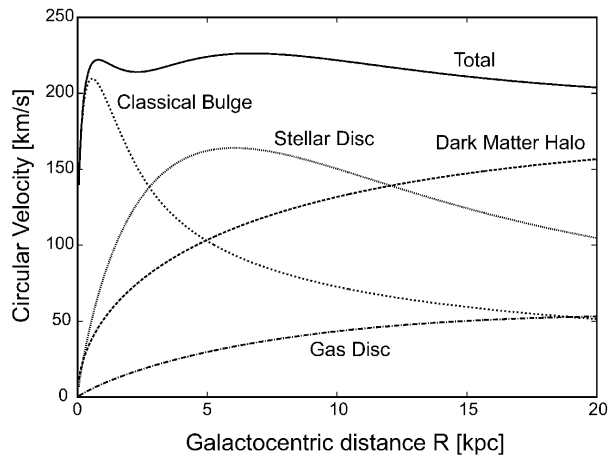}
\caption{
Initial circular velocity curves of each component of the galaxy.
}	
\label{fig:InitRC}
\end{center}
\end{figure}

We generated the initial axisymmetric model of a stellar and gaseous discs with a classical bulge embedded in a DM halo. 
Fig. \ref{fig:InitRC} shows the initial circular velocity curves of each component in the model.

The stellar disc follows an exponential profile: 
\begin{eqnarray}
 \rho_{\rm disc}(R,z) = \frac{M_{d}}{4\pi R_{d}^2 z_{d}} \exp\left(-\frac{R}{R_{d}}\right){\rm sech}^2\left(\frac{z}{z_{d}}\right), 
\end{eqnarray}
where $M_{d}$, $R_{d}$, and $z_{d}$ are the total mass, scale-length, and scale-height of the stellar disc, respectively.
Using Hernquist's method \citep{Hernquist1993}, the velocity structure of the stellar disc in cylindrical coordinates 
is approximately determined by a Maxwellian approximation.
We set the reference radial velocity dispersion by assuming 
that Toomre's $Q$ at $R=2.5 R_d$ equals $1.3$ \citep[see section 2.2.3 of][]{Hernquist1993}.

The gas disc also follows an exponential profile with a total mass of $M_{d,g}$, scale-length of $R_{d,g}$, 
and scale-height of $z_{d,g}$. The gas disc was truncated at $5 R_{d,g}$.
The initial temperature was set to $10^4$ K, and the initial velocity dispersion was $10~\rm km~s^{-1}$ everywhere.

The classical bulge follows the Hernquist profile:
\begin{eqnarray}
 \rho_{\rm bulge}(r) = \frac{M_{b,0}}{2\pi}\frac{a_b}{r(r+a_b)^3},
\end{eqnarray}
where $M_{b,0}$ and $a_b$ are the total mass and scale-length of the bulge \citep{Hernquist1990}, respectively.
Following \citet{WidrowDubinski2005}, we generated the classical bulge from the distribution function
with an energy cutoff \citep[see Equation (10) of][]{WidrowDubinski2005}. Here, we set $q_b$ of that equation to be 0.21.

For simplicity, we assumed the DM halo to be a static potential, whose density profile follows 
the Navarro-Frenk-White profile:
\begin{eqnarray}
 \rho_{\rm halo}(r) =  \frac{M_{h}}{4\pi f_c(C_{h})}\frac{1}{r(r+a_h)^2}, \\
 f_c(x) = \ln(1+x) - x/(1+x), 
\end{eqnarray}
where $M_{h}$, $a_{h}$, and $C_{h}$ are the total mass, virial radius, and 
concentration parameter of the of the DM halo, respectively \citep{Navarro+1997}, and $x \equiv r/a_h$.
As shown in Fig. \ref{fig:InitRC}, the contribution from the stellar disc to the total circular velocity in the galaxy model 
surpasses that from the DM haloes in the regions of $3 < R < 12$ kpc. 
In this case, the galaxy satisfies the criteria for the bar instability \citep{Efstathiou+1982} 
and spontaneously develops a stellar bar.
Note that a static halo omits dynamical friction on the bar \citep[e.g.][]{DebattistaSellwood2000,AthanassoulaMisiriotis2002}; 
however, dynamical friction will be weak because the central density of the DM halo in this study is low. 
We will investigate effects of a `live' DM halo on the dynamical evolution of barred spirals in a future study.

The initial numbers of stars and SPH particles were $6.4$ million and $4.5$ million, respectively, 
and the particle masses of stars and SPH particles were 
about $9.1 \times 10^3~M_{\rm \odot}$ and $3 \times 10^3 ~M_{\rm \odot}$, respectively. 
We used a gravitational softening length of $10$ pc.
Following the above procedures, we generated a galaxy model in a near-equilibrium state 
and then subsequently allowed the galaxy to evolve for $\sim 10 - 20$ rotational periods under 
the constraint of axisymmetry \citep{McMillanDehnen2007,Fujii+2011}.
This equilibrium state was used as the initial condition for the numerical simulation.

\subsection{Quantification of spiral arms}
\label{sec:Analysis}

The quantities that characterize the spiral structure are the number of spiral arms, 
their shape (i.e. logarithmic, hyperbolic, segments of lines), and their amplitude (or arm-to-inter-arm contrast). 
In order to analyse these quantities, we used the one-dimensional (1D) Fourier decomposition method 
with respect to the azimuthal direction \citep[see][]{RixZaritsky1995}. 
More specifically, we divided the stellar surface density $\mu(R,\phi)$ into strips using a polar coordinates $(R,\phi)$ 
and then performed a 1D Fourier decomposition of the each strip with respect to the azimuthal direction:
\begin{eqnarray}
B_m(R) = \int_{-\pi}^{\pi} \frac{\mu(R,\phi)}{\bar{\mu}(R)} \exp[-im\phi] d\phi,
\label{eq:FFT}
\end{eqnarray}
where 
$\bar{\mu}(R)$ is the azimuthally averaged surface stellar density at radius $R$, and 
$B_m(R)$ denotes a complex amplitude of the $m$-th mode.

Radial profiles of the pitch angle were calculated using radial changes of phases of $B_m(R)$.
If we define the phase as $\phi_m(R) = \arctan(B_m/|B_m|)$, then we can evaluate the pitch angle at radius $R$ 
using the following equation:
\begin{eqnarray}
 \cot i_{{\rm p},m} (R) = R \frac{d\phi_m(R)}{dR}.
\end{eqnarray}

We analysed the angular speeds of the spiral arms with the Fourier density peak method.
In this method, the rate of angular phase changes of $B_m$, $\Omega_{{\rm phase},m}(R,t)$, 
is referred to the `angular phase speed' \citep[][]{Wada+2011,Baba+2013,Roca-Fabrega+2013}.

\section{Results and Discussion}
\label{sec:results}

\begin{figure*}
\begin{center}
\includegraphics[width=0.98\textwidth]{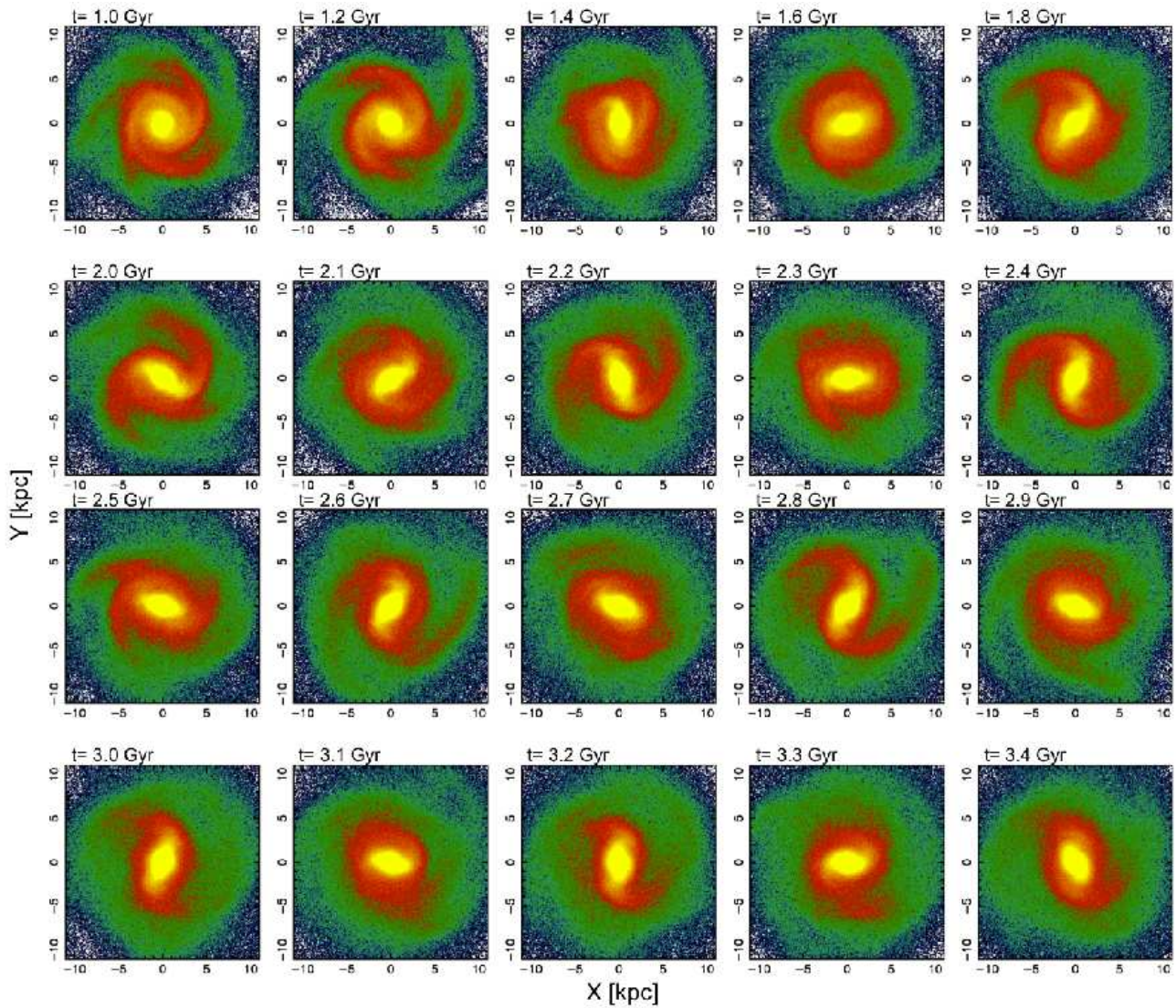}
\caption{
Face-on stellar mass distributions of the galaxy every 200 Myr from 1.0 Gyr to 2.0 Gyr,
and every 100 Myr from 2.0 Gyr to 3.4 Gyr.
The bar instability occurs around $t \sim 1.3$ Gyr. After the bar is well developed,
grand-design spirals are clear around 2.0, 2.2, 2.4, 2.6, 2.8, 3.0, 3.2, and 3.4 Gyr.
}	
\label{fig:SnapshotEvolutionEvery100Myr}
\end{center}
\end{figure*}

We present the results of the dynamical evolution of barred grand-design spirals.
Fig. \ref{fig:SnapshotEvolutionEvery100Myr} shows the time evolution of the stellar mass distribution over a period of 1.0 -- 3.4 Gyr. 
The simulated galaxy has an $m=2$ grand-design spiral without a stellar bar until $t \simeq 1.2$ Gyr, 
begins to show bar instabilities around $t \simeq 1.3$ Gyr, and then reaches a quasi-steady state until $t \simeq 2$ Gyr.
As shown in previous numerical simulations, we confirm that the grand-design spiral is neither steady nor always connected with the stellar bar.
For example, the $m=2$ grand-design spiral is clear at $t = 2.0$ Gyr, becomes faint at $t=2.1$ Gyr, and appears clearly again at $t=2.2$ Gyr. 
This unambiguously shows that grand-design spirals change their strengths with a period of about 200 Myr.

In the following subsections, we first describe the short-term (i.e. $\lesssim 200$ Myr) behaviour of grand-design spirals 
and compare these results with the predictions from the swing amplification and invariant manifold theories.

\subsection{Short-term behaviour of grand-design spirals}
\label{subsec:shortterm}

\begin{figure*}
\begin{center}
\includegraphics[width=0.76\textwidth]{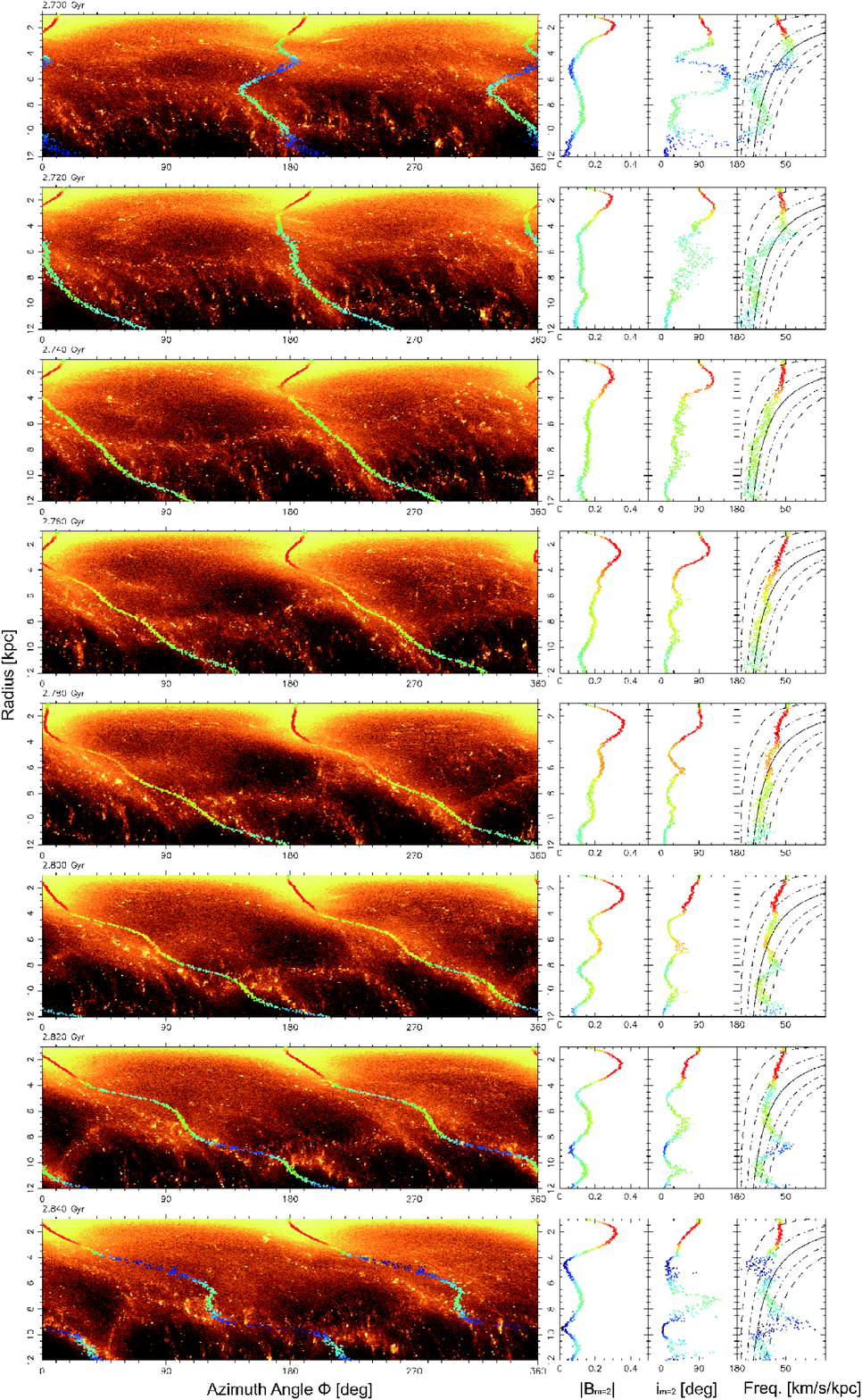}
\caption{
Evolution of the spiral arms and the $m=2$ mode every $20$ Myr.
The first column shows the distribution of the stars in polar coordinates (i.e. $R$--$\phi$ plane). 
The peak phases of the $m=2$ mode are indicated by dotted lines whose colours (rainbow) 
correspond to the $m=2$ mode amplitudes.
Time evolution is shown in a rotating frame of the bar with approximately $47~\rm km~s^{-1}~kpc^{-1}$. 
The horizontal axis indicates the azimuthal angle $\phi$ [degrees], 
and the vertical axis denotes the galactocentric distance $R~\rm [kpc]$.
The remaining three columns (left to right) plot the amplitude $|B_{m=2}|$, pitch angle $i_{m=2}$ [degrees], 
and pattern speed $\Omega_{{\rm phase},m=2}$ [$\rm km~s^{-1}~kpc^{-1}$], respectively.
}
\label{fig:SpiralEvol}
\end{center}
\end{figure*}

The left panels in Fig. \ref{fig:SpiralEvol} shows the time evolution of the spiral arms in the polar coordinates $(R,\phi)$ from 2.70 -- 2.84 Gyr. 
Thisspan corresponds to one cycle of a bar-spiral connection event. 
During this term, the bar and strong grand-design spiral appear at $t \simeq$ 2.76 -- 2.78 Gyr.
If we focus on the grand-design arm placed at $\phi = 180^\circ$ -- $270^\circ$, 
we see that this arm is initially short and weak ($t=2.7$ Gyr) 
and then grows into a grand-design spiral arm with winding until $t=2.74$ Gyr.

The right panels in Fig. \ref{fig:SpiralEvol} show the time evolution of radial profiles of the amplitude ($|B_{m=2}|$), 
pitch angle ($i_{m=2}$), and angular phase speed ($\Omega_{{\rm phase},m=2}$) of the $m=2$ mode.
We observe that, at the {\it clear} grand-design spiral phase (i.e. $t \simeq$ 2.76 -- 2.78 Gyr), 
the spiral parameters are nearly constant with respect to the radius or slowly decrease from 3 kpc to the outer regions.
In contrast, at the damping phase ($t \gtrsim 2.8$ Gyr), the radial profiles of all spiral parameters show radial modulations.
Note that the radially changing pitch angle at the clear barred-spiral phase seems to contradict the traditional picture, 
where the observed geometries of spiral arms are frequently fitted by a logarithmic spiral 
\citep[i.e. constant pitch angle;][]{Kennicutt1981,KennicuttHodge1982}.
However, as per recent measurements of observed spiral galaxies, the pitch angle for most galaxies 
decreases as the distance from the centre increases \citep[e.g.][]{SavchenkoReshetnikov2013},
which is consistent with our results. 

\begin{figure*}
\begin{center}
\includegraphics[width=0.95\textwidth]{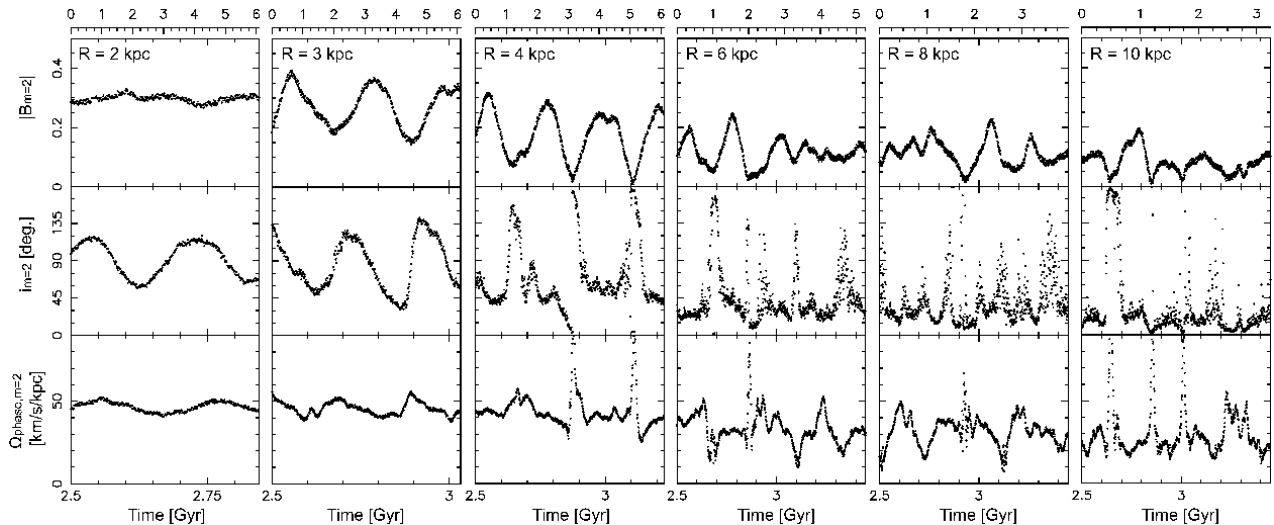}
\caption{
Evolution of the spiral parameters at each radius (width of 1 kpc) from 2.5 Gyr. 
The upper horizontal axes shows the time normalized by the rotational period $T_{\rm rot}$ at each radius.
The spikes in evolution of pitch angle are caused by ill-determined Fourier modes 
when the amplitudes are very small.
}
\label{fig:SpiralParameters}
\end{center}
\end{figure*}

The top row in Fig. \ref{fig:SpiralParameters} shows the time evolution of the amplitudes at each radius.
As described above, the amplitudes change with a period of about 200 Myr at each radius,
although the amplitude at $R = 2$ kpc (i.e. the bar region) is more stable than those of the other regions.
The upper horizontal axis in each panel measures the time normalized by the rotational period $T_{\rm rot}$ at each radius.
We see that the amplitude at each radius changes in time spans of 1 -- 2 $T_{\rm rot}$;
that is, the grand-design spirals in barred spiral galaxies are not quasi-stationary patterns, 
but transient recurrent structures (i.e. dynamic spirals)\footnote{
For unbarred spiral galaxies, \citet{SellwoodCarlberg2014} argued that the dynamically evolving spiral patterns are 
attributed to the superposition of a few longer-lived waves, each of which has well-defined frequencies and shapes
and survives for 5 -- 10 rotations. 
}.

\begin{figure*}
\begin{center}
\includegraphics[width=0.95\textwidth]{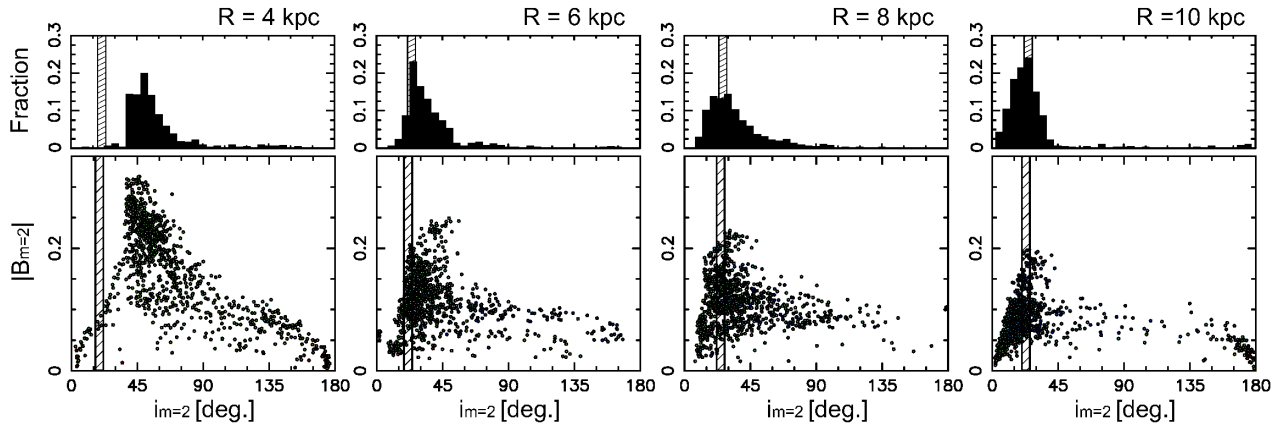}
\caption{
Bottom: Evolution of the $m=2$ modes at $R =$ 4, 6, 8, and 10 kpc on the $i_{m=2}$ --  $|B_{m=2}|$ plane. 
Each dot represents a value sampled every 1 Myr.
Top: Amplitude-weighted frequencies of the appearance of pitch angle $i_{m=2}$ at each radius.
The hatched region corresponds to the predicted maximum pitch angle $i_{\rm swa}$ 
around the analysed region due to the swing amplification (see Section \ref{subsec:SWA}).
}
\label{fig:PitchAngleAmplitudePlane}
\end{center}
\end{figure*}

As shown in the middle row in Fig. \ref{fig:SpiralParameters}, the pitch angles also change similarly, 
but their phases are inversely correlated with the phases of the amplitude changes.
In other words, spiral arms become strong as they wind from open to closed 
and then become damped after reaching the maximum amplitude.
Such evolutionary behaviour between $|B_{m=2}|$ and $i_{m=2}$ is clearly shown 
in the bottom panels in Fig. \ref{fig:PitchAngleAmplitudePlane}, which 
shows the values of $|B_{m=2}|$ and $i_{m=2}$ at $R =$ 4, 6, 8, and 10 kpc every 1 Myr.
We see that, at all radii, the amplitudes are very small in leading spirals ($i_{m=2} > 90^\circ$) 
and tightly winding trailing spirals ($i_{m=2} \lesssim 10^\circ$),
while the amplitudes are large in open trailing spirals ($i_{m=2} \simeq 20^\circ$ -- $45^\circ$).

\begin{figure*}
\begin{center}
\includegraphics[width=0.95\textwidth]{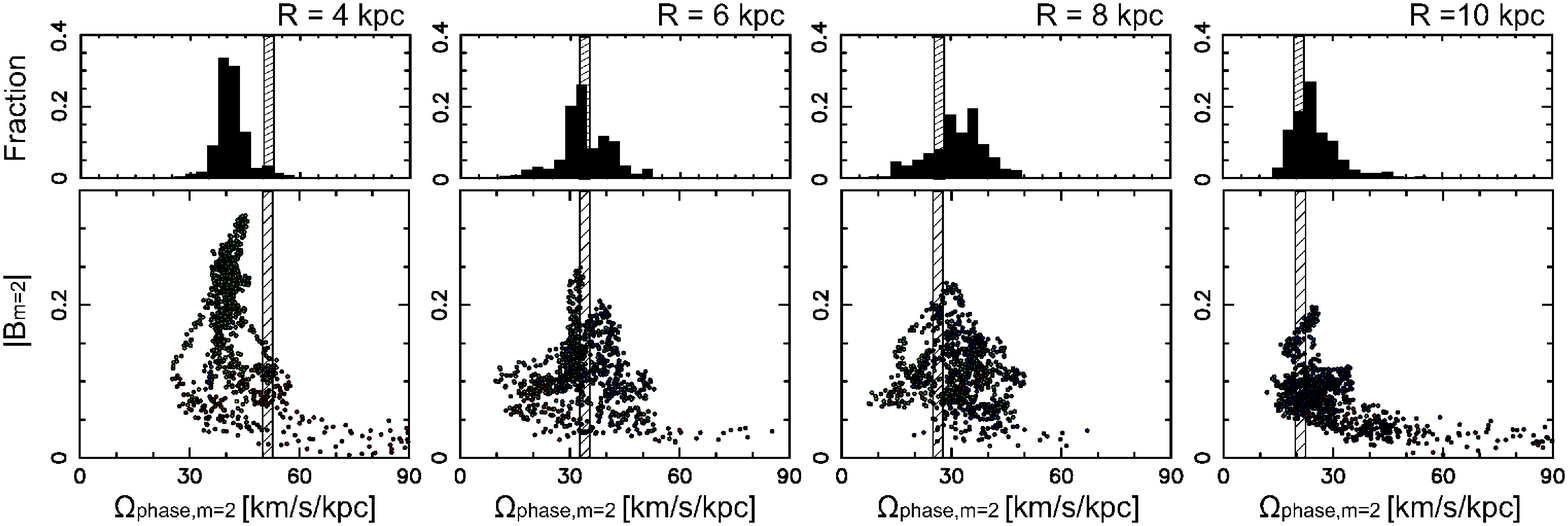}
\caption{
Bottom: Evolution of the $m=2$ modes at $R =$ 4, 6, 8, and 10 kpc on $|B_{m=2}|$--$\Omega_{{\rm phase},{m=2}}$ plane. 
Each dot represents a value sampled every 1 Myr.
Top: Amplitude-weighted frequencies of the appearance of phase pattern speed $\Omega_{{\rm phase},m=2}$ at each radius.
The hatched region indicates the angular speed of the galaxy $\Omega(R)$ at each radius.
Note that the bar's pattern speed (at $R \lesssim 3$ kpc) is approximately $47~\rm km~s^{-1}~kpc^{-1}$.
}
\label{fig:AmplitudePatternSpeedPlane}
\end{center}
\end{figure*}

The bottom panels in Fig. \ref{fig:SpiralParameters} show 
the time evolutions of the phase speeds, $\Omega_{{\rm phase},m=2}$, at each radius.
The phase speeds at $R \lesssim 3$ kpc are stable and approximately $45~\rm km~s^{-1}~kpc^{-1}$
because the bar dominates in these regions. 
In contrast, the phase speeds both increase and decrease compared to the galactic rotation for all radial ranges,
and as for the evolution of the pitch angles, the phases are inversely correlated to the phases of the amplitude changes.
The bottom panels in Fig. \ref{fig:AmplitudePatternSpeedPlane} clearly show 
such evolutionary behaviour between the phase speed and amplitude. 
When the phase speed at each radius is approximately equal to the galactic angular speed, $\Omega(R)$,
the amplitude reaches a maximum.

Finally, we discuss the evolution of the bar. As shown in the left panels in Fig. \ref{fig:SpiralEvol}, 
the bar with a semi-major axis of about 3 kpc is always located around $\phi = 0$ and 180$^\circ$,
even though its parameters show the time evolution. 
The pitch angles and phase speeds of the right panel in Fig. \ref{fig:SpiralEvol} 
show flat radial profiles at $R \lesssim 2$ kpc. 
In particular, at the clear grand-design spiral phase (i.e. $t \simeq$ 2.76 -- 2.78 Gyr), 
we can see that these flat profiles reach $R \simeq 3$ kpc.
As shown in Fig. \ref{fig:SpiralParameters}, the amplitude and angular phase speed at $R = 2$ kpc 
are relatively stable, suggesting that the inner structure of the bar is steady.
However, the amplitude and angular phase speed clearly change in time at $R = 3$ kpc.
Such behaviour of the stellar bar is usually seen in $N$-body simulations of 
isolated discs \citep{Martinez-Valpuesta+2006,Dubinski+2009} 
as well as cosmological simulations of disc galaxies \citep{Okamoto+2014}.

\subsection{Comparison with swing amplification theory}
\label{subsec:SWA}

We compare our results with the swing amplification theory 
to investigate the physical origin of the short-term evolutional behaviour among spiral parameters.
Before this comparison, we briefly review the physical essence of the swing amplification
\citep[see][and references therein for details]{DobbsBaba2014}.
The swing amplification operates through a combination of the shearing flow, epicyclic motions, and disc self-gravity,
which play roles in promoting or suppressing the formation of structures in disc galaxies \citep{Toomre1981}.
Since the direction of epicyclic motion of a star is the same as the direction 
which the spiral arm is sheared by differential rotation, stabilisation by rotation is reduced,
and the perturbation can grow via the usual Jeans instability \citep{GoldreichLynden-Bell1965,JulianToomre1966}. 
As a concrete example, we consider a disc galaxy with a nearly flat rotation curve
as our simulated galaxy (see Fig. \ref{fig:InitRC}).
Because the angular speed follows as $\Omega \propto R^{-1}$, 
the timescale of epicyclic motion is given by 
\begin{eqnarray}
 \tau_{\rm epi} \equiv \frac{1}{\kappa} = \left(R\frac{d\Omega^2}{dR}+4\Omega^2\right)^{-1/2}= \frac{1}{\sqrt{2}\Omega},
\end{eqnarray}
while the timescale of involvement with the spiral arm is given by
\begin{eqnarray}
 \tau_{\rm shear} \equiv \frac{1}{R |d\Omega/dR|} = \frac{1}{\Omega}.
\end{eqnarray}
Comparing the above two time-scales, we can see that $\tau_{\rm epi} \simeq \tau_{\rm shear}$.
The shearing motion cancels out the stabilisation effect of epicyclic motions,
and thus, the structure can grow via self-gravity in a short time as compared to $\tau_{\rm epi}$.

To compare the results described in Section \ref{subsec:shortterm} with the predictions from the swing amplification theory, 
we calculate the pitch angle at the maximum amplitude phase, $i_{\rm SWA}$, expected from the swing amplification theory. 
Following \citet{MichikoshiKokubo2014}, $i_{\rm SWA}$ is given by 
\begin{eqnarray}
  i_{\rm SWA} = \arctan\left(\frac{2}{7}\frac{\sqrt{4-2\Gamma}}{\Gamma}\right),
\end{eqnarray}
where $\Gamma = -d\ln\Omega/d\ln R$ is a dimensionless shear rate. 
This equation predicts that $i_{\rm SWA}$ is a decreasing function of $\Gamma$.
This trend matches observations \citep[e.g.][]{Seigar+2006}
and $N$-body simulations of disc galaxies \citep{Grand+2013}.

In bottom panels in Fig. \ref{fig:PitchAngleAmplitudePlane}, we plot the expected values of $i_{\rm SWA}$ as shaded regions.
In the outer regions at  $R=8$ kpc and $10$ kpc, the peak pitch angles (around $25^\circ$--$30^\circ$) 
of the simulations match the predictions well. 
Similar results have already been obtained for $N$-body simulations of multiple arm spiral galaxies \citep{Baba+2013}.
In contrast, the peak pitch angles at the inner regions (i.e. $R=4$ kpc and $6$ kpc) are 
somewhat larger than the predicted values due to the effect of the central bar.
Furthermore, we also compare the frequency of appearance for $i_{\rm m=2}$ with $i_{\rm SWA}$.
The top panels in Fig. \ref{fig:PitchAngleAmplitudePlane} show that 
the most frequent pitch angle matches the values predicted by the swing amplification theory, except for $R=4$ kpc.

In addition, the swing amplification theory can explain the evolutionary behaviour between $\Omega_{{\rm phase},m=2}$ and $|B_{m=2}|$
because the amplification is most efficient around the co-rotating radius \citep[see][and references therein]{DobbsBaba2014}.
In fact, the top panels in Fig. \ref{fig:AmplitudePatternSpeedPlane} show that 
the most frequent phase speed decreases as a function of the radius, 
but it is the same as or slightly faster than the galactic angular speed, except for $R=4$ kpc.
The previous $N$-body simulations of barred spiral galaxies have also pointed out that
the spiral arms rotate with a decreasing angular rotation speed
that is slightly faster than the galactic angular speed \citep{Grand+2012b}.
However, these studies have not been compared with the swing amplification theory.

\subsection{Effects of the bar on spiral dynamics} 
\label{subsec:barspiralinteraction}

In the above subsections, we presented the short-term behaviour of the grand-design spiral 
and showed that these results can be explained by the swing amplification mechanism.
However, this mechanism cannot explain the behaviour in the inner regions (e.g. $R \lesssim 4$ kpc). 
In this subsection, to clarify behaviour of the spiral at the inner regions, 
we investigate how the bar affects the dynamics of the grand-design spirals.

\begin{figure*}
\begin{center}
\includegraphics[width=0.95\textwidth]{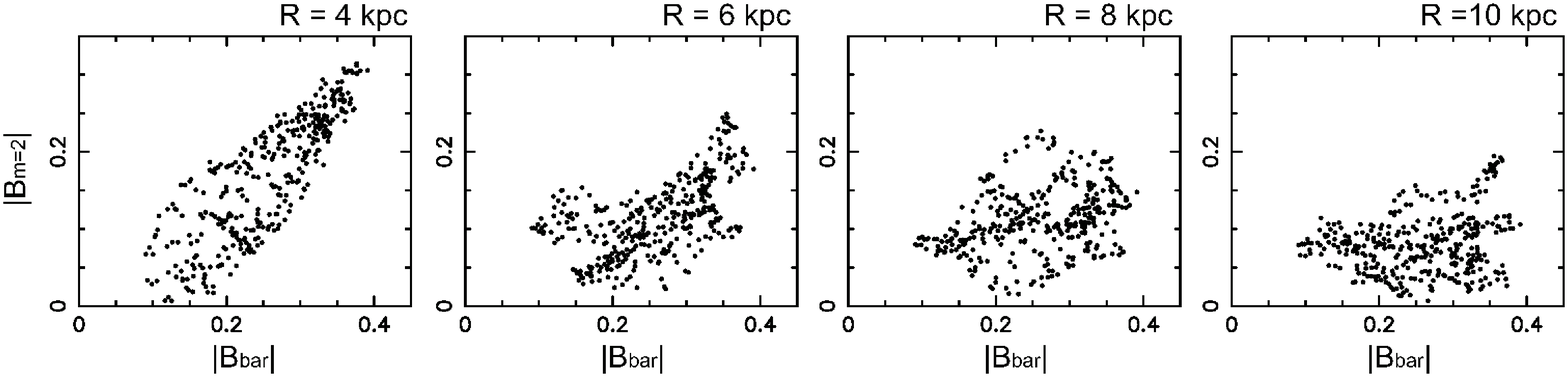}
\caption{
Bar amplitude $|B_{\rm bar}|$ with respect to the spiral amplitude $|B_{m=2}|$ at $R =$ 4, 6, 8, and 10 kpc.
The bar amplitude $|B_{\rm bar}|$ is measured with the amplitude of the $m=2$ mode at $R=3$ kpc.
Each dot represents a value sampled every 1 Myr.
}
\label{fig:BarSpiralAmplitude}
\end{center}
\end{figure*}

To clarify the cone of influence of the bar, we plot the results for the (spiral) amplitude $|B_{m=2}|$
with respect to the bar amplitude $|B_{\rm bar}|$ in Fig. \ref{fig:BarSpiralAmplitude}.
Here, we measure $|B_{\rm bar}|$ with the amplitude of the $m=2$ mode at $R=3$ kpc.
At $R =$ 4 kpc, there is a clear trend showing that, as  the spiral amplitudes become larger, the bar becomes stronger. 
We see a similar trend at $R =$ 6, but it has considerable scatter around the correlation.
In contrast, the results at $R=8$ and 10 kpc show no clear correlation.
These results suggest that the bar affects the dynamics of the spiral arm 
well beyond its co-rotation radius ($\simeq 4$ kpc; see Fig. \ref{fig:SnapshotEvolutionEvery100Myr}) 
and possibly $\lesssim 6$ kpc. This maximum radius corresponds to approximately 1.5--2 bar radius 
and is consistent with observational values \citep[approximately 1.4--1.6;][]{Salo+2010}.
We, therefore, suggest that the bar drives the spiral arm within approximately 1.5--2 bar radius.

\begin{figure*}
\begin{center}
\includegraphics[width=0.95\textwidth]{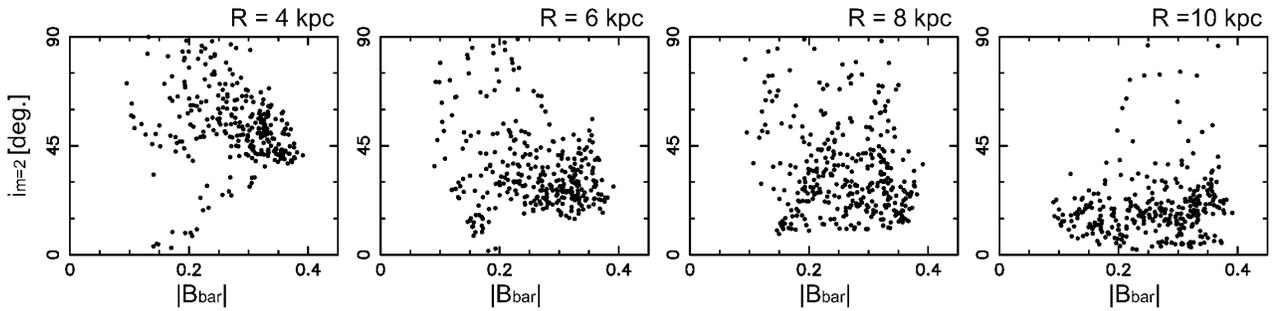}
\caption{
Bar amplitude $|B_{\rm bar}|$ and spiral pitch angle $i_{m=2}$ at $R =$ 4, 6, 8, and 10 kpc.
Each dot represents a value sampled every 1 Myr.
}
\label{fig:BarSpiralPitchAngle}
\end{center}
\end{figure*}

\begin{figure*}
\begin{center}
\includegraphics[width=0.950\textwidth]{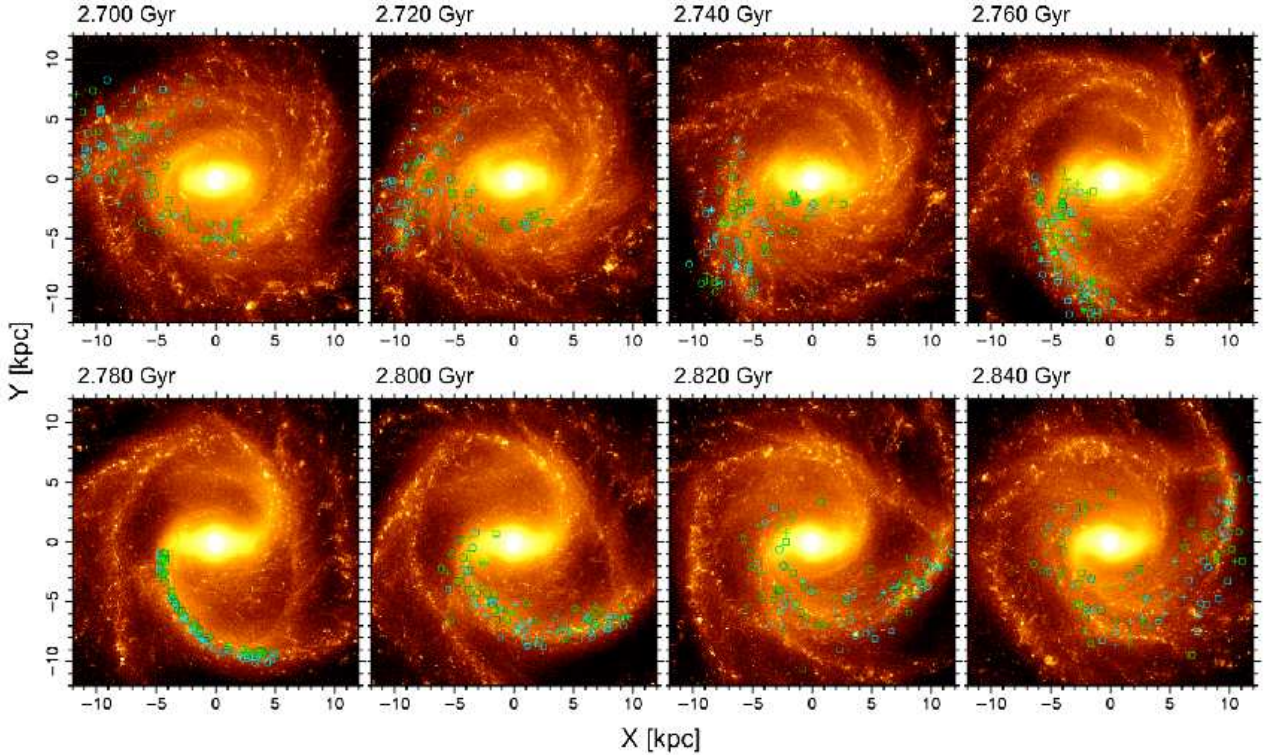}
\caption{
Motion of selected star particles with respect to the rotating frame of the stellar bar. 
The star particles are selected from the particles associated with a grand-design spiral at $t=2.78$ Gyr.
Time evolution is shown in a rotating frame of the bar with approximately $47~\rm km~s^{-1}~kpc^{-1}$. 
}	
\label{fig:StarTracer}
\end{center}
\end{figure*}

To see the dynamical behaviour of the bar-driven spirals, 
we investigate whether the inner spiral arms are consistent with the predictions from the invariant manifold theory. 
As described in Section \ref{sec:intro}, the manifold theory provides two predictions: 
(1) the spirals in strongly barred galaxies will be more open than those in less strongly barred ones, and
(2) the spiral arms should be a bundle of orbits guided by the manifolds 
so that stars should move along the arms rather than across them.
We compare the simulated barred spirals with these predictions. 
Fig. \ref{fig:BarSpiralPitchAngle} shows the results for the spiral pitch angle $i_{m=2}$ versus $|B_{\rm bar}|$.
We cannot see any correlation between $i_{m=2}$ and $|B_{\rm bar}|$ for $R=8$ and 10 kpc.
In contrast, for $R = 4$ and 6 kpc, we see that the low pitch angle spirals ($i_{m=2} \lesssim 20^\circ$) 
are lacking for stronger bars ($|B_{\rm bar}| \gtrsim 0.2$). 
In this sense, our results correspond to the predictions from the invariant manifold theory,
although there is considerable scatter for weaker bars.
Furthermore, to see whether stars move along the spiral arm as predicted from the invariant manifold theory,
we select the star particles located at the grand-design spiral at $t=2.78$ Gyr
and then follow the trajectories of these particles from 2.70 Gyr to 2.84 Gyr.
These trajectories are shown in Fig. \ref{fig:StarTracer}.
At $t<2.76$ Gyr, the chosen stars are located at both sides of the arm, and these stars form the arm at $t \simeq 2.78$ Gyr.
After that, these stars escape from the arm to both sides. 
Therefore, the motions of the stars therefore are not simply consistent with the prediction from the invariant manifold theory. 

\begin{figure*}
\begin{center}
\includegraphics[width=0.90\textwidth]{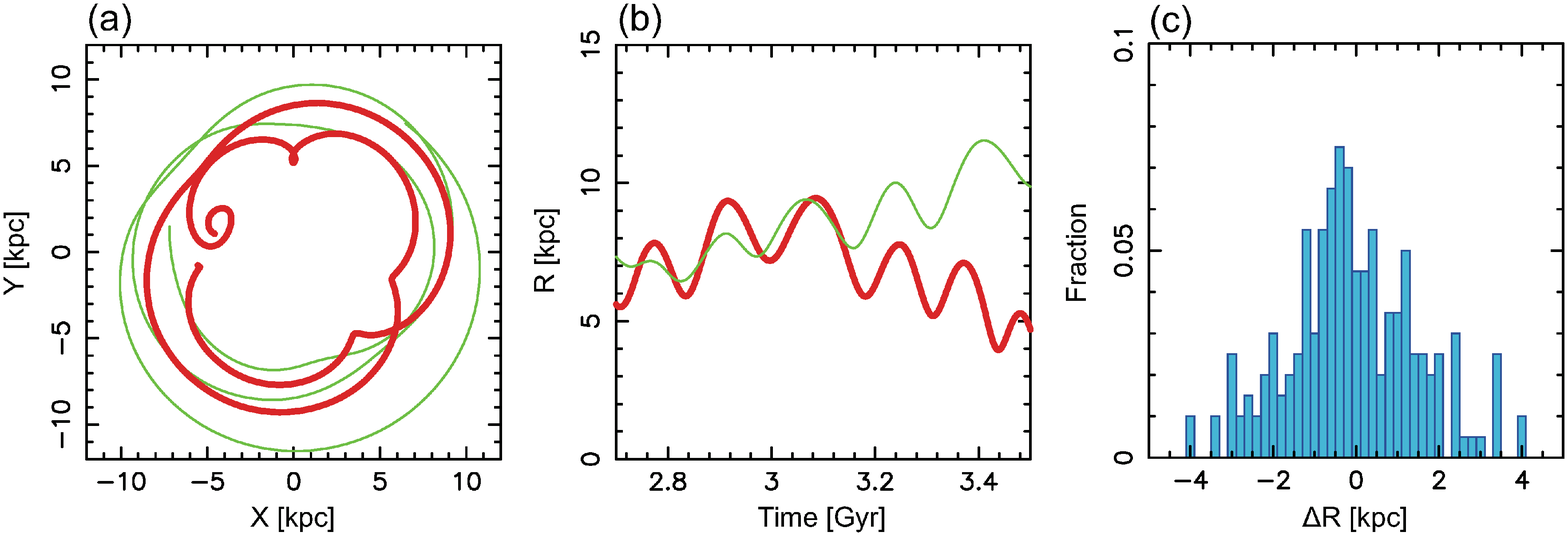}
\caption{
(a) Sample of the orbits of the stars shown in Fig. \ref{fig:StarTracer}.
Time evolution is shown in a rotating frame of the bar at approximately $47~\rm km~s^{-1}~kpc^{-1}$. 
(b) Time evolution of the radii of the stars shown in panel (a).
(c) Probability distribution of the radial migration $\Delta R$ [kpc] of the stars shown in Fig. \ref{fig:StarTracer}.
Here, $\Delta R \equiv {\bar R}_2 - {\bar R}_1$, where ${\bar R}_1$ and ${\bar R}_2$ are the time-averaged radii 
of stellar orbits at $2.5 < t<2.8$ Gyr and $3.0 < t < 3.3$ Gyr, respectively.
}	
\label{fig:Orbits}
\end{center}
\end{figure*}

\citet{Struck2015} suggested that co-rotating spirals with bars 
are supported by resonant orbits excited in outer Lindblad zones, which are the parents of a continuous sequence of resonant radii. 
To investigate whether this is true, we plotted sample orbits of these particles in the rotating frame in Figs. \ref{fig:Orbits}a and \ref{fig:Orbits}b. 
We could not find any resonant closed orbits, rather most orbits show monotonic centre shifts of the radial oscillations 
(i.e. radial migrations of stars; Fig. \ref{fig:Orbits}c) due to scattering by the spiral arms as presented by 
$N$-body simulations of galactic discs \citep{SellwoodBinney2002,Roskar+2012,Grand+2012b,Baba+2013,Grand+2014}. 
This result suggests that the  {\it collective motion} of stars might be an essential ingredient for generating dynamic spirals due to 
a non-linear combination among the shearing flow, non-linear epicyclic (or p-ellipse) motions, and self-gravity (see Section \ref{subsec:SWA}). 
However, this orbit analysis is limited to the stars selected from a single snapshot. 
Therefore, it would be interesting to further explore the orbits of stars comprising dynamic spirals in numerical simulations 
for comparison with kinematic theories of stars to aid bar and spiral formation \citep{Romero-Gomez2012,Struck2015}.

\section{Conclusions and Implications}
\label{sec:summary}

In this study, we performed a 3D$N$-body/SPH simulation of a barred spiral galaxy with 
our original $N$-body/SPH code {\tt ASURA-2} \citep{SaitohMakino2009,SaitohMakino2010} 
and investigated the short-term evolution of grand-design spirals in barred galaxies. 
We summarize our main results and give implications below.

\begin{enumerate}
\item As suggested by previous simulations \citep[e.g.][]{SellwoodSparke1988,RautiainenSalo1999}, 
{\it clear} grand-design spirals in barred galaxies appear {\it only when} the spirals connect with the ends of the bar. 
This implies that barred galaxies with spirals {\it not} starting from bar ends 
will have weaker spirals than those in barred galaxies with spirals starting from the bar ends.  
Actually, some barred spiral galaxies exhibit a clear bar--spiral phase differences \citep[e.g. NGC 3124 and NGC 3450;][]{Buta+2007} 
and their spirals are not clear grand-design spirals. Yet, this is not a statistical argument. 
To confirm our implication, it requires a statistical comparison between 
barred spirals with and without the bar--spiral phase difference. 

\item Grand-design spiral arms in a barred galaxy are not rigidly rotating patterns 
but rather differentially rotating dynamic patterns, as suggested by previous simulations \citep{Baba+2009,Grand+2012b}. 
This means that the amplitudes, pitch angles, and phase speeds are not constant, 
but change within a time span of 1--2 rotational periods at each radius. 
Such behaviour suggests that the assumption of rigidly rotating grand-design spirals is not necessarily reasonable 
for investigating gas flows and cloud formation in barred spiral galaxies.
This point will be discussed further in Baba, Morokuma-Matsui \& Saitoh (\citeyear{Baba+2015b})
and a forthcoming paper (Baba et al. in preparation).

\item The stellar bar affects the evolution of spiral arms within 1.5--2 bar radius,
and this radius is consistent with observations of barred spiral galaxies \citep{Salo+2010}. 
Within this radius, as predicted from the invariant manifold theory \citep[e.g.][]{Athanassoula+2009a}, 
low pitch angle spirals are lacking for stronger bars, 
although there is considerable scatter for weaker bars in the bar amplitude--spiral pitch angle plane. 
In addition, motions of stars located at a grand-design spiral are not simply consistent with 
the prediction from the invariant manifold theory. 

\item The swing amplification theory can explain the most frequent pitch angles and 
the amplitude--pitch angle relationships in the outer regions (i.e. $R >$ 1.5--2 bar radius). 
These results imply that interplay among the galactic shearing motion, epicyclic motion of stars, 
and self-gravity of stars is an essential factor for the dynamics of spiral arms even in barred spiral galaxies. 
Note that we assumed that the DM halo is a fixed external potential, 
so that the bar is formed via bar instabilities in massive discs \citep{OstrikerPeebles1973,Efstathiou+1982}. 
In this case, the self-gravity of the disc plays a role in the formation of spiral arms via the swing amplification. 
Considering a `live' DM halo, however, \citet{Athanassoula2002} showed that 
bars can develop even in less massive discs. 
We infer, then, that the spiral arms in such discs may show different behaviour 
from those in massive discs such as those investigated in this study. 
This topic will be the subject of future work. 

\end{enumerate}

\bigskip
\section*{ACKNOWLEDGEMENTS}

I would like to thank the referee, Curtis Struck, for many constructive comments and suggestions 
which have significantly improved the paper.
I thank Keiichi Wada, Michiko S. Fujii, Jun Kumamoto, Daisuke Kawata, 
and Kana Morokuma-Matsui for a careful reading of the manuscript.
I also thank Takayuki R. Saitoh for technical supports on performing numerical simulations with {\tt ASURA-2}.
Calculations, numerical analyses and visualization were carried out on Cray XT4, XC30,
and computers at Center for Computational Astrophysics, National Astronomical Observatory of Japan.  
This research was supported by HPCI Strategic Program Field 5 `The origin of matter and the universe'
and JSPS Grant-in-Aid for Young Scientists (B) Grand Number 26800099.


\end{document}